\documentclass[12pt]{iopart}
\usepackage{setstack}

\usepackage{graphicx}
\usepackage[T1]{fontenc}
\usepackage{cite}
\usepackage{units}
\usepackage{color}

\newcommand{\beq}{\begin{equation}}
\newcommand{\eeq}{\end{equation}}
\begin{document}
\title[Nanoscale sensing based on nitrogen vacancy centers]{Nanoscale sensing based on nitrogen vacancy centers in single crystal diamond and nanodiamonds: achievements and challenges}

\author{M Radtke$^1$, E Bernardi$^1$\footnote{Present address:
Istituto Nazionale di Ricerca Metrologica, Strada delle cacce 91, 10137 Torino (To), Italy},  A Slablab$^1$, R Nelz$^1$, and  E Neu$^1$} 

\address{$^1$ Faculty for natural sciences and technology, Physics, Saarland University, 66123 Saarbr\"ucken, Germany}

\begin{abstract}
Powered by the mutual developments in instrumentation, materials and theoretical descriptions, sensing and imaging capabilities of quantum emitters in solids have significantly increased in the past two decades. Quantum emitters in solids, whose properties resemble those of atoms and ions, provide alternative ways to probing natural and artificial nanoscopic systems with minimum disturbance and ultimate spatial resolution. Among those emerging quantum emitters, the nitrogen-vacancy (NV) color center in diamond is an outstanding example due to its intrinsic properties at room temperature (highly-luminescent, photo-stable, biocompatible, highly-coherent spin states). This review article summarizes recent advances and achievements in using NV centers within nano- and single crystal diamonds in sensing and imaging. We also highlight prevalent challenges and material aspects for different types of diamond and outline the main parameters to consider when using color centers as sensors. As a novel sensing resource, we highlight the properties of NV centers as light emitting electrical dipoles and their coupling to other nanoscale dipoles e.g. graphene.  
\end{abstract}

\maketitle

\section{Introduction}
During the past two decades, color centers in diamond emerged as sensors for various relevant quantities like electric, magnetic and optical near fields as well as local temperature and strain \cite{Dolde2014, Rondin2014, Tisler2013a, Kucsko2013, Teissier2014}. As point defects with tightly-localized electrons, color centers in diamond resemble atomic-scale quantum systems. Thus, color centers open up routes towards nanoscale sensing technologies in which the spatial resolution is no longer limited by the sensor's volume \cite{Chernobrod2005}. However, ultimate spatial resolution will only be attainable using individual defect centers. This approach simultaneously limits the intensity of the photoluminescence (PL) light which is mainly being used as the sensor read-out. Consequently, sacrificing ultimate spatial resolution, color center ensembles can provide enhanced sensitivity, speed up measurements and lower experimental complexity.  In addition to being atomic-scale sensors, color centers are highly-sensitive, potentially quantum-enhanced sensors providing coherent superpositions of spin states as sensing resource. Diamond as a host material has a multitude of outstanding characteristics ranging from high optical transparency, to mechanical hardness and chemical inertness. Diamond also has the rare property that its crystal lattice is naturally almost free of nuclear spins as the most abundant carbon isotope (${}^{12}$C, 98.9\%, \cite{Balasubramanian2009}) has nuclear spin zero. Consequently, the lattice has low magnetic noise (given that the concentration of paramagnetic impurities like e.g.\ nitrogen is low) and electronic spins of color centers can have long coherence times reaching the millisecond range even at room temperature. However, diamond's high refractive index renders light extraction from color centers challenging and often demands structuring diamond to yield photonic structures or to use nanodiamonds (NDs). Due to this fact, the fields of diamond photonics and nanofabrication are strongly connected to sensing using color centers.  

In 1997, the first observation of spin manipulation of an individual negatively-charged nitrogen vacancy (NV) center was reported \cite{Gruber1997}. In 2000, stable, room-temperature single photon emission from individual NVs added to their potential for quantum technologies \cite{Kurtsiefer2000}. The NV defect consists of a nitrogen atom replacing a carbon atom and a vacancy on a neighboring lattice site. For simplicity, NV center refers to the negatively-charged defect throughout this review.  Since then, other defects have been characterized as single emitters. Among these novel defects there are several group IV-based centers namely silicon vacancy (SiV) \cite{Becker2017review}, germanium vacancy (GeV) \cite{Siyushev2017}, tin vacancy (SnV) \cite{Iwasaki2017} and even lead-related defects \cite{Tchernij2018Lead}. In contrast to NV centers that show an almost $\unit[100]{nm}$  broad emission band, group IV-based defects concentrate almost all PL into their only a few nanometer wide zero phonon lines (ZPLs) even at room temperature. Temperature dependent shifts of these ZPLs have been exploited for all optical temperature sensing (see e.g.\ Refs.\ \cite{Fan2018,Ngyuen2018}). Despite recent progress in sensing using group IV defects, negatively-charged NVs remain outstanding in the field of sensing. Their superior usability for sensing applications stems from the fact that the NV center's electronic spins can be optically polarized using convenient, non-resonant excitation at room temperature while they can be manipulated using microwave radiation in an easily accessible frequency band (\unit[2.88]{GHz}). Internal population dynamics of the NV center lead to different PL intensities for different spin states. Consequently, the spin state can be read out using this PL difference (optically-detected magnetic resonance, ODMR). However, sensing using NV centers is not limited to using the spin degree of freedom: Potential sensing resources include the NV's charge state, modulations of excited state lifetimes and absorption. The latter two exploiting the dipolar nature of the NV center's transitions.
  
In this paper, we will review recent advances in sensing with NV centers. We will summarize main breakthroughs and  drawbacks in the field. The review is structured as follows: Section \ref{sec_diamondmat} presents basics of diamond material synthesis and structuring for sensing applications. Section \ref{sec_NVsensingmain} starts with a short introduction of NV properties and then details the main parameters characterizing NV sensors. In Section \ref{sec_reso}, we summarize the spatial resolution attainable with NV sensors in different sensing modes. Section \ref{sec_applications} summarizes selected applications to illustrate recent progress using NV sensors.     
 
\section{Manufacturing and structuring  of nano- and single crystal diamonds \label{sec_diamondmat}}
\subsection{Manufacturing of diamond \label{sec:diamondgrowth}}
\begin{figure}
\centering
\includegraphics[width=10 cm]{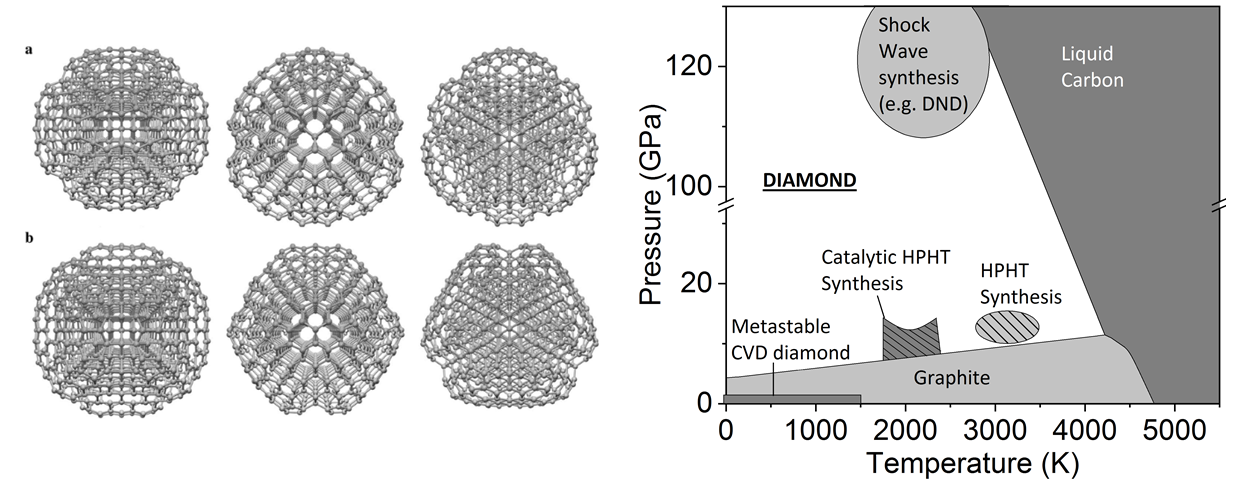}
\caption{left: Structures of nanodiamond supercells from various crystallographic axes. Reproduced with permission from \cite{lai_surface_2012}. right: Diamond phase diagram based on \cite{bundy_p_1980}  \label{fig:diamondcellsphase}}
\end{figure}

The major factors influencing diamond's properties and its potential applications are purity and crystallinity. The following retrospective summarizes recent developments within diamond growth and synthesis according to those parameters. 
In theory, diamond is a crystal consisting only of sp$^{3}$ hybridized covalently bound carbon atoms, which build a repetitive unit cell made of fused hexagonal chair cyclohexane conformations as depicted in Fig.\ \ref{fig:diamondcellsphase} \cite{dahl_isolation_2003}. The Hermann Mauguin notation classifies diamond into the Fd3m-07h face centered cubic group. The unique arrangement of carbon atoms within diamond makes it not only the hardest material in Mohs scale, but also causes it to be inert towards chemical modifications without use of harsh conditions. Diamond's wide bandgap (indirect gap \unit[5.45]{eV}) results in optical transparency for light up to $\approx$\unit[230]{nm}. The naturally almost spin free lattice of diamond (${}^{12}$C, 98.9\%, I=0, \cite{Balasubramanian2009}) has been engineered by isotopically purifying diamond (residual ${}^{13}$C, 0.3\%) as presented in Ref.\ \cite{Balasubramanian2009}.

\begin{figure}
\centering
\includegraphics[width=15 cm]{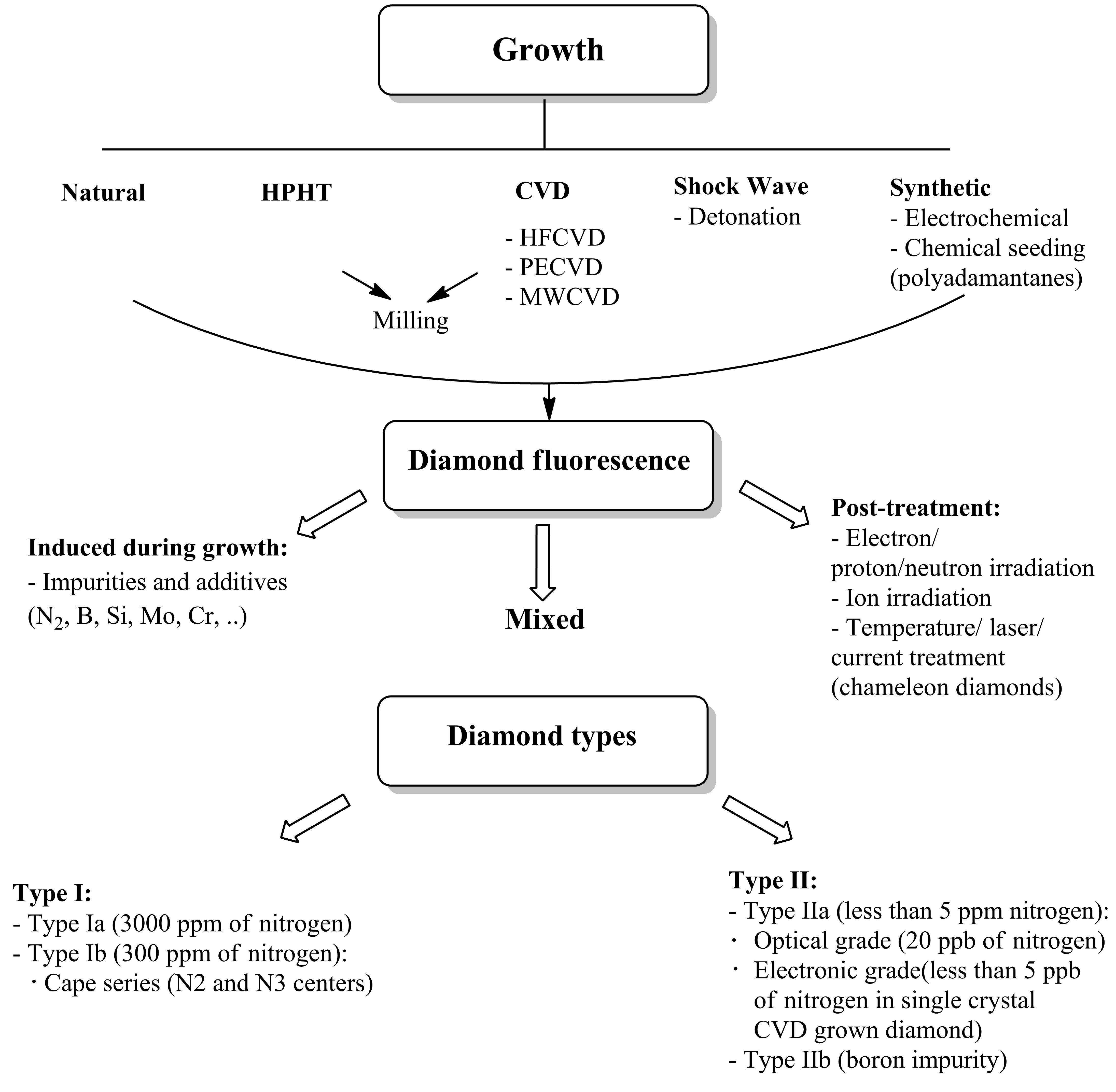}
\caption{Classification of diamond, details see text. \label{fig:classification}}
\end{figure}

There are many types of diamonds and their classification is based on several factors. One classification system has its foundation in the origin of the diamonds: naturally occurring diamonds, diamond created via synthetic hydrothermal high-pressure high temperature (HPHT) protocols, shock wave synthesis (e.g.\ through detonation) and low pressure pyrolysis of hydrocarbon gases in chemical vapor deposition (CVD). The latter includes microwave assisted CVD (MWCVD) and hot filament CVD (HFCVD) deposition, whereas MWCVD is most suitable to grow high purity diamonds \cite{Markham2010}. Fig.\ \ref{fig:classification} summarizes this classification. For CVD diamond growth, the most common precursor gases are methane CH$_{4}$ and hydrogen H$_{2}$. The detonation method only forms nanodiamonds (NDs) of about \unit[5]{nm} size \cite{Krueger2007}. 
Figure \ref{fig:diamondcellsphase} summarizes the temperature/pressure regimes for diamond synthesis through shock wave, metastable CVD diamond phase generation or HPHT protocols. In both HPHT and CVD synthesis, by tailoring the additives in the reaction chamber (e.g.\ insertion of N$_{2}$), diamonds with various color centers are obtained. For detonation diamonds (DNDs), creating luminescent NDs is challenging. However, recently density control of NV centers in \unit[5]{nm}-sized DNDs was reported  \cite{sotoma_enrichment_2018}. Usually DNDs require special post-treatment to remove graphitic sp$^{2}$ layers, e.g.\ through immersion in a boiling mixture of oxidizing mineral acids (HNO$_{3}$, H$_{2}$SO$_{4}$ and HClO$_{4}$) and thermal annealing in vacuum or under oxygen for adequate stabilization of spectral properties \cite{fang_preparation_2018}. Various synthesis methods generate diamond with tailored properties. For example DNDs are commonly used as abrasives in industry, while HPHT and CVD (especially single crystal) diamonds are often more suitable for quantum applications. 

For HPHT and CVD methods, diamond growth is typically initiated on a substrate or seed. For CVD growth, this can be attained by directly growing on a diamond (homoepitaxy) or via using a non-diamond substrate (heteroepitaxy) which needs to be prepared suitably.  Substrate pretreatments for CVD span from seeding with NDs to grow polycrystalline diamond to creation of nuclei using bombardment with carbon ions (bias enhanced nucleation for heteroepitaxy) \cite{Schreck2017}. Homoepitaxy limits the size of grown SCDs due to the size of available substrates. This can be overcome using heteroepitaxy \cite{Schreck2017}. 

The smallest possible seeds for diamond growth are so called diamondoids. Diamondoids mainly combine one or more adamantane molecules (C$_{10}$H$_{16}$), the smallest unit cage structure of the diamond crystal lattice. Using the smallest possible seed is among other reasons motivated by minimizing the influence of the seed on the resulting ND's properties. Refs.\ \cite{dahl_isolation_2003,balaban_nmr_2015} demonstrate the synthesis of nanoscale 'diamond molecules' (higher diamondoids, e.g.\ tetramantanes, less than \unit[1]{nm} in size). Diamondoids were recently employed to enable the growth of high quality NDs in a HPHT process \cite{alkahtani2019} and in a CVD process \cite{tzeng2017vertical}. Ref.\ \cite{gebbie_experimental_2018} reports that diamondoids serve as nuclei in CVD if they contain more than 26 carbon atoms. Another approach to molecular sized diamonds is the use of meteoritic diamonds \cite{Vlasov2014}. For this material class, fluorescent SiV centers have been observed in NDs with only \unit[1.6]{nm} size. However, this results has never been obtained using man-made NDs, thus significantly lowering applicability.   
A novel route to manufacturing an extreme diamond nanomaterial is the recently observed transformation of a bilayer of graphene into a single layer of (fluorine terminated) diamond \cite{bakharev_chemically_2019}. 

The second system that classifies diamond, in addition to the origin of the diamond, is based on the crystallinity. Here diamonds are classified as single crystal (SCD), polycrystalline and (ultra-)nanoscrystalline diamonds according to their crystallite size. For grain sizes below 10 nm the diamond is typically called ultrananocrystalline, between \unit[10]{nm} and \unit[50]{nm} nanocrystalline, and from \unit[50]{nm} to \unit[500]{$\mu$m}  micro- or polycrystalline. If the crystallite size exceeds \unit[500]{$\mu$m}, the diamond is termed single crystalline and a specified crystallographic growth plane can be assigned \cite{may_ultrananocrystalline_2008}.

\begin{figure}
\centering
\includegraphics[width=0.6 \textwidth]{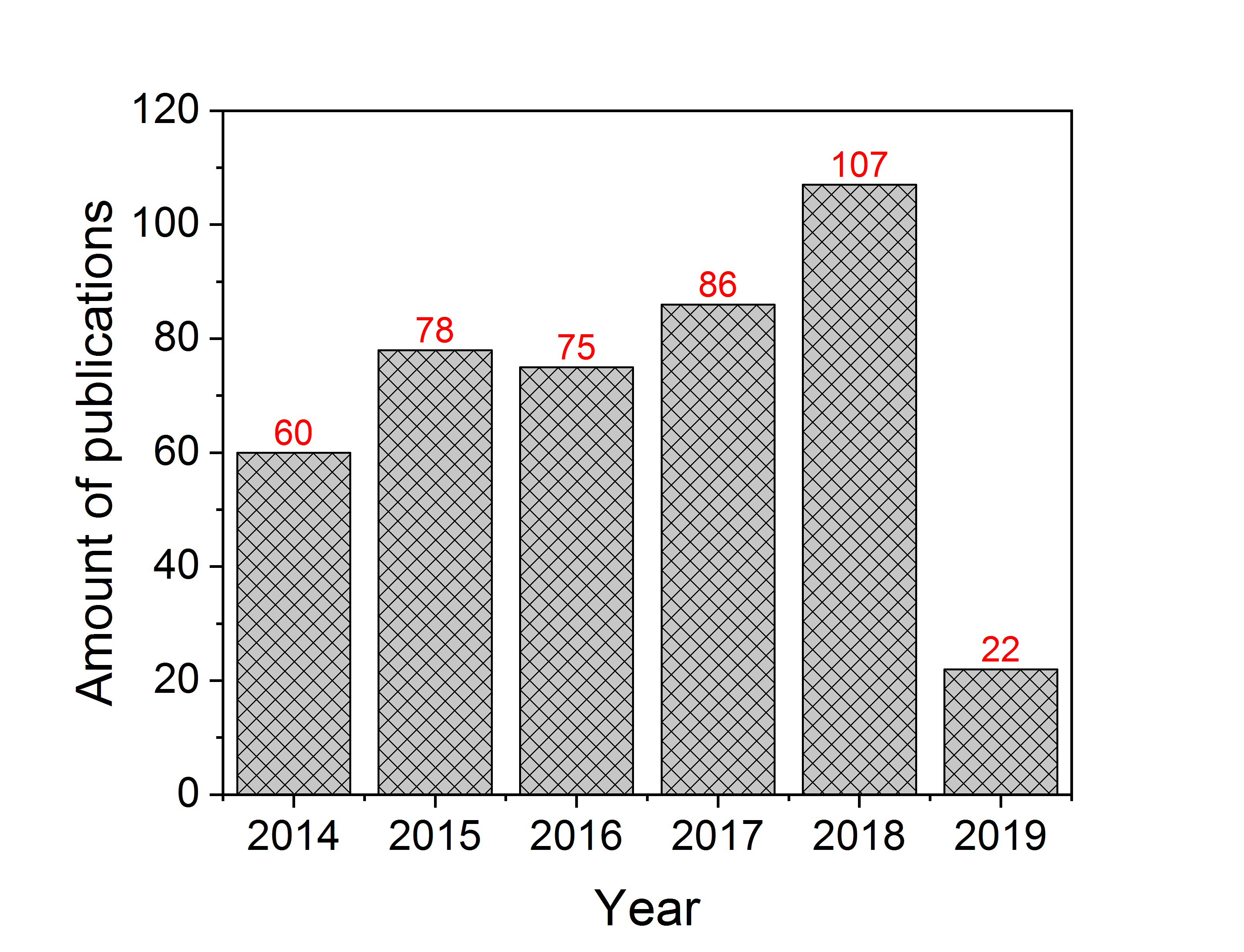}
\caption{Timescale in period between 2013-2018 showing the evolution of the searching phrase "single crystal diamond growth", based on the database WorldCat.org.  \label{fig:worldcat}}
\end{figure}

The amount of publications detailing the phrase 'single crystal diamond growth' in WorldCat.org. in the time period between 2013-2018 yields 3.575 results (Fig.\ \ref{fig:worldcat}), illustrating the significance of the field. 

Free standing wafer-sized polycrystalline diamond films are currently commercially available with diameter up to \unit[140]{mm} (Element Six, UK). Recently, Ref.\ \cite{Schreck2017} presented a mm-thick SCD wafer with \unit[98]{mm}  diameter grown by heteroepitaxy. The  basic usability of this material for sensing with NV centers has been demonstrated \cite{Nelz2019}.

The most common growth direction in CVD is (100). However, for many applications it is advantageous to control the orientation of NV centers with respect to the SCD surface. Thus, alternative growth of (111) \cite{Tallaire2014} diamond and (113) \cite{Lesik2015} has been realized that orients the dipoles of NV centers more optimal with respect to photonic structures manufactured into the SCD surface \cite{Neu2014}. While single SCD allows for the highest purity material with the lowest background PL, poly- and nanocrystalline forms will always suffer from broadband PL occurring at grain boundaries \cite{Markham2010}.

The third classification has been historically established for natural diamonds and classifies them according to their optical absorption \cite{Robertson1934}. Figure \ref{fig:classification} summarizes this classification. Type IIa diamonds were classified by a lack of optical absorption \cite{Robertson1934}, so this were the purest diamonds in this classifications. The absorption of type I diamond is nowadays assigned to nitrogen \cite{Walker1979}. Type Ia diamonds contain nitrogen in aggregated form; while in type Ib diamonds, it is present in its isolated form substituting single carbon atoms (substitutional nitrogen) \cite{Walker1979}. Most CVD diamonds are of IIa type, while HPHT diamonds are often type Ib \cite{Balmer2009}. In contrast, type I diamonds are known to occur most commonly in nature. Type IIb contains boron in small quantities and has p-doped character. This historical classification doesn't sufficiently classify diamonds typically used for sensing with color centers: These diamonds are mostly so pure that they are all class IIa but need further classification. So typically, the amount of substitutional nitrogen and boron is directly given. Pushing purity to the limit, electronic grade diamonds  with less than \unit[5]{ppb} nitrogen and less than \unit[1]{ppb} boron are nowadays commercially available from ElementSix and are the basis for most sensing experiment especially using individual NV centers. In addition, $^{12}$C isotopically-purified diamond  has been demonstrated and termed quantum grade diamond \cite{Balasubramanian2009}.  

When impurities in diamond are present in high concentration, they induce a coloration of diamonds. For sensing, typically low concentrations of defects are used, which does not necessarily lead to a coloration of diamond.

Color centers can be introduced into diamond using two approaches. The first approach creates color centers in the growing diamond via using selected precursors \cite{Ohno2014,Ishiwata2017,Michl2014}. Often NV centers are also created due to background nitrogen in the process \cite{Neu2014, Lesik2014, Doi2016, Michl2014} . However, for sensing applications it is important to carefully control color center density and depth below the diamond surface (see also Sec.\ \ref{sec_reso}). Creating thin layers containing NV centers has been attained in diamond CVD using $\delta$-doping (e.g.\ Ref.\ \cite{Ohno2014, ohno2012engineering}): in a CVD process with low growth rate, nitrogen gas is introduced for a short time thus inducing an only \unit[6]{nm} thick nitrogen-doped layer \cite{Ohno2014}. Moreover, nitrogen doped layers were obtained by overgrowing nitrogen-terminated diamond surfaces \cite{Chandran2016}. Using $\delta$-doping, conversion of nitrogen impurities to NV centers is often fostered via irradiation e.g.\ using helium ions \cite{deOliveira2016}. One advantage of forming color centers directly during diamond growth is that they can show preferential alignment: instead of aligning along all equivalent directions in the crystal, color centers only form with one orientation leading to a perfectly aligned ensemble of centers (see Refs.\ \cite{Lesik2014,Ishiwata2017,Michl2014}). Moreover, NV centers created using conversion of grown-in nitrogen to NV centers have been found to have superior properties compared to centers arising due to implanted nitrogen \cite{ruf2019optically,vanDam2019}.

The second approach creates impurities after diamond manufacturing has finished. It utilizes irradiation (by electrons, neutrons or protons \cite{wolf2015subpicotesla,ruf2019optically}) and ion implantation \cite{de2017tailoring,lovchinsky2016nuclear,tetienne2018spin, yamamoto2013extending,Radko2016,broadway2018bandbending}. It allows to control the fluence, which determines the density and the energy (ranging from keV to MeV) to set the implantation depth \cite{prins_ion_2003}. The creation of color centers can be also controlled in lateral dimension by implanting through a pierced AFM tip \cite{riedrich-moller_nanoimplantation_2015}, using nanomasks \cite{jakobi_efficient_2016} or focused ion beams \cite{Lesik2013}.
Ion implantation in diamond  requires  post-treatment annealing (mostly at temperatures exceeding 600$^{\circ}$C) in order to induce vacancy migration, repair implantation damage and form color center complexes like NV centers.  

\subsection{Nanostructuring of diamond \label{sec:nanostructuring}}

\begin{figure}
\centering
\includegraphics[width=15 cm]{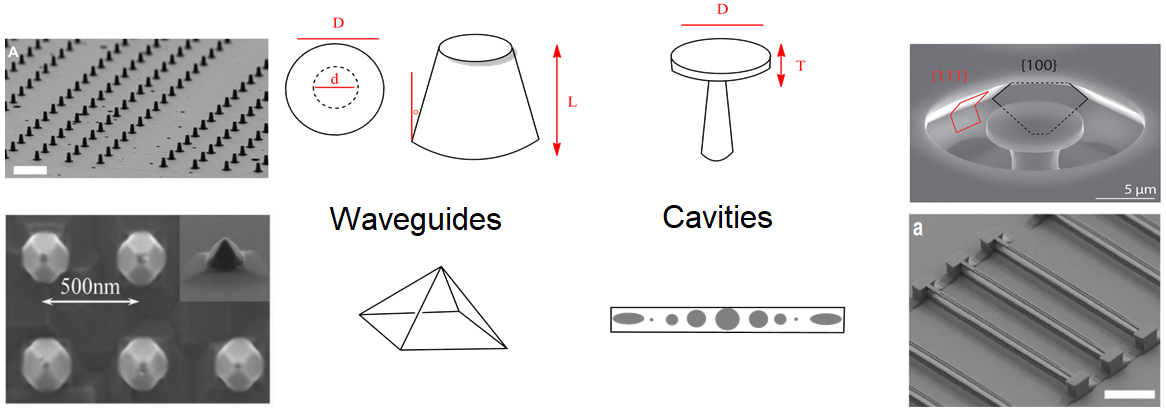}
\caption{Current state of the art examples of successful nanofabrication performed on diamond as an example of nanowaveguides and nanocavities. Reproduced with permission from \cite{momenzadeh_nanoengineered_2015, mitchell_realizing_2019, mitchell_realizing_2019, burek_high_2014, jaffe_deterministic_2019}. \label{Fig:nanostructures}}
\end{figure}

To efficiently extract color center PL from diamond, photonic structures e.g.\ nanopillar waveguides \cite{Marseglia2018, Neu2014}, optical antennas \cite{Riedel2014}, solid immersion lenses \cite{Wildanger2012}, bulls eye gratings \cite{li2015efficient} or pyramids \cite{jaffe_deterministic_2019,Nelz2016,choi2018enhancing} are highly-desirable (for reviews on diamond photonics, see Refs.\ \cite{Beha2012a,Aharonovich2014a,Schroder2016,Atature2018}). Figure \ref{Fig:nanostructures} displays some examples of diamond nanostructures. Nanopillar waveguides channel the light into conveniently collectible emission angles (NA <0.95) and circumvent total internal reflection \cite{Babinec2010}. Here, the high refractive index of diamond (2.4) aids in confining the light in the waveguide. In contrast, for unstructured diamond, it limits the collection efficiency to only a few percent \cite{Fuchs2018}. While optical antennas, in principle, allow up to unity collection efficiency \cite{Riedel2014}, tip-like structures as pillars or pyramids might simultaneously serve as a tip for a scanning probe microscope \cite{Appel2016} thus enabling to scan a color center in close proximity to a sample under investigation. It is noteworthy that nanopillars have been produced incorporating not only NV centers \cite{Babinec2010,mccloskey2019enhanced,Neu2014} but also SiV centers \cite{Marseglia2018} and lead-related defects \cite{Trusheim2019}. Additionally, cavity structures like one or two-dimensional photonic crystals \cite{burek_high_2014,riedrich-moller_nanoimplantation_2015} or microdisk (whispering gallery mode) resonators \cite{mitchell_realizing_2019} allow the coupling of color centers to cavity modes with high Q-factor. Mostly, photonic structures are used to extract light from single color centers \cite{Marseglia2018}. However, very recent studies also indicate the usefulness of pillar arrays for sensing with NV ensembles \cite{mccloskey2019enhanced}. 

To harness the full potential of photonic structures, nanofabrication processes must be carefully controlled: First, the shape of the nanostructure will strongly influence its performance e.g.\ changing the taper angle of a nanopillar [see Fig.\ \ref{Fig:nanostructures}] changes the efficiency of light extraction \cite{momenzadeh_nanoengineered_2015,Fuchs2018}. Second, nanofabrication, especially on example of plasma etching, might damage diamond surfaces \cite{Kato2017}.  This damage was estimated to reach several nanometers into the diamond and found to deteriorate the properties of shallow color centers \cite{Oliveira2015}. Reduced damage and enhanced properties of NV centers have been reported by using plasmas with zero bias voltage and thus minimal acceleration of ions in the plasma \cite{Oliveira2015} and with careful combinations of etching processes \cite{ruf2019optically}. As diamond is highly chemically inert, wet chemical etching methods are not an alternative to plasma-based processing.  Focused ion beam (FIB) milling leaves gallium impurities in the diamond that need to be removed using sophisticated post-processing \cite{Riedrichmoeller2011}. So the most common processing techniques utilize standard nano-lithography techniques with electron beams, laser or 2-photon lithography (2PP) for mask generation and transfer the patterns into diamond using reactive ion etching (RIE), mostly in the form of inductively coupled reactive ion etching (ICP-RIE) \cite{ortiz-huerta_fabrication_2018-1,sun_controlled_2016}. Plasma-etching of diamond requires special gases as well as resists for the masks (negative and positive tone). Most commonly oxygen/argon or chlorine-based mixtures are used. Masks are either based on PMMA with hard mask techniques (Al, Al$_{2}$O$_{3}$, W) or directly on the negative tone resist hydrogensilsequioxane (HSQ) \cite{Neu2014,tao_single-crystal_2014}. 

Nanostructuring of diamond can be classified into traditional nanotechnology approaches: bottom up and top down approaches. For example nanodisc resonators with quality factor exceeding 300000 may be generated by top-down techniques involving electron-beam lithography, laser lithography and ICP-RIE \cite{mitchell_realizing_2019}.

In bottom-up approaches, nanostructures are directly formed during diamond growth thus potentially forming many structures at the same time and avoiding potential damage due to plasma-etching \cite{Nelz2016, Aharonovich2013}. However, top-down approaches have the advantage, that color centers can be created while the diamond is still in its bulk form, which eases annealing and cleaning followed by fabrication of the nanostructures (e.g.\ Ref.\ \cite{Appel2016}).

\section{Achievements, improvements and challenges in NV sensing \label{sec_NVsensingmain}}
The strengths of sensing using color centers in diamond are its spatial resolution and its sensitivity (for a tutorial on NV magnetometry see e.g.\ Ref.\ \cite{abe2018tutorial}). Moreover, depending on the application, it is important to consider in which modality the diamond sensor should be operated. This especially includes the host diamond's geometry (bulk, single-crystal diamond (SCD), nanostructures or nanodiamonds (NDs)) and how the color center is brought into interaction with the sample (e.g.\ NDs entering cells, SCD tip scanning over the sample, sample transferred to SCD surface). Moreover, a decision needs to be rendered if single color centers should be used or if ensembles are more suitable. Ensembles can provide better sensitivity whereas single centers, in combination with scanning probe operation, provide ultimate spatial resolution.
In the following, we give a very short introduction to the physics of NV centers. We are not aiming for completeness nor we will describe any theoretical background here but just introduce the basic level scheme and effects needed to follow the discussed sensing topics. For details on the NV center see e.g.\ Ref.\ \cite{Manson2013}. 
   
\begin{figure}
\begin{center}
  \includegraphics[width=0.8\textwidth]{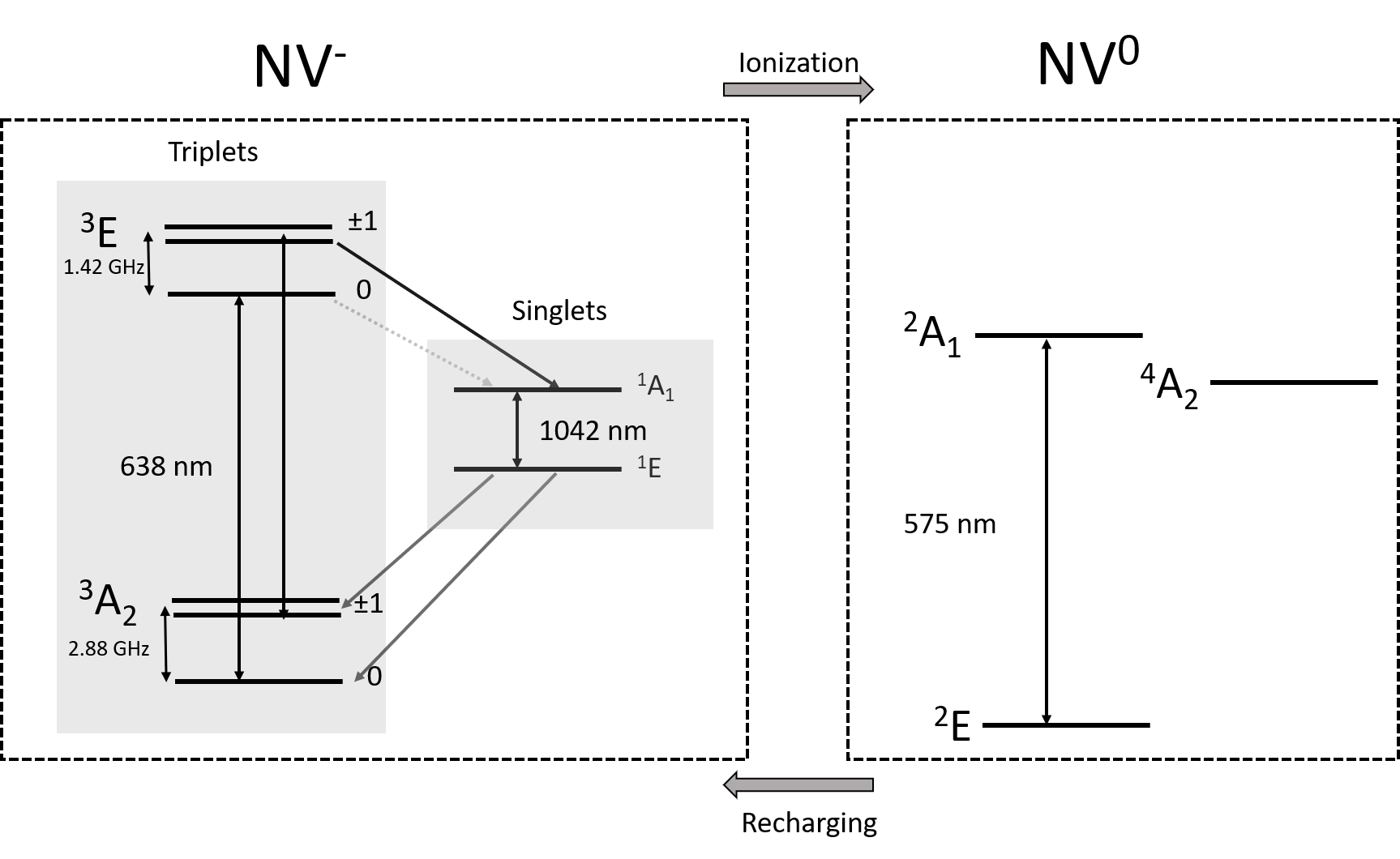}
  \caption{Basic level scheme of the negatively-charged and the neutral NV center. Explanation see text.}
  \label{fig:NVlevelscheme}
	\end{center}
\end{figure}
\subsection{NV Basics \label{sec:NVbasics}}
The NV center consists of a nitrogen atom that substitutes a carbon atom in the diamond lattice and a neighboring vacancy. Negatively-charged NV centers display a system of singlet and triplet levels. A strong optical transition (lifetime \unit[12.9]{ns}, \cite{Collins1983}) occurs between the ${}^{3}$A$_2$ ground state and the ${}^{3}$E excited state. This transition leads to the zero-phonon-line (ZPL) at \unit[637]{nm} at room temperature. Due to phonon assisted transitions, the narrow ZPL is accompanied by a broad phonon sideband spanning almost \unit[100]{nm} spectral width \cite{Kehayias2013}.  The ${}^{3}$A$_2$ $\to$ ${}^{3}$E transition, which induces the visible photoluminescence (PL) of the NV center, is represented by two orthogonal transition dipoles. These dipoles are placed in the plane orthogonal to the NV's high-symmetry axis that connects the nitrogen and the adjacent vacancy. This axis is oriented along one of the equivalent [111]-directions of the diamond lattice \cite{davies1976optical, epstein2005anisotropic}.

In the triplet ground state of NV$^-$, three spin states with m$_s$=-1, 0 and +1 are observed. Between the m$_s$=0 level and the degenerate m$_s$=-1,+1 levels a zero-field splitting of 2.87 GHz occurs.  These levels show spin-dependent PL: A NV center in the m$_s$=0 cycles between ${}^{3}$A$_2$ and ${}^{3}$E and produces bright PL. In contrast, a NV center in the m$_s$=-1 or +1 state undergoes an intersystem crossing to the singlet states with higher probability. As the center will reside in these states for typically around 200 ns \cite{Robledo2011a}, the visible PL of the center is reduced. The PL difference between the two spin states under continuous, non-resonant laser excitation (e.g.\ using \unit[532]{nm} laser light) can be up to 20\% \cite{Rondin2014}. Simultaneously, non-resonant laser excitation polarizes the NV center into the m$_s$=0 state. Following this initialization, transitions between the spin levels can be driven using microwaves.

Between their ${}^{1}$A$_1$ and ${}^{1}$E levels, NV centers display a transition in the infrared spectral range (\unit[1042]{nm}) \cite{Kehayias2013}. As a result of non-radiative processes, the PL due to this transition is typically four orders of magnitude weaker than the visible PL \cite{Rogers2008}. Nevertheless, transitions in the singlet system are a valuable resource for sensing (see Sec.\ \ref{sec:sensitivity}).

If one electron is removed from the NV$^-$ center, it converts to a neutral NV$^0$ center. This occurs during optical excitation of NV$^-$  when a NV$^-$ center which resides in the excited state absorbs an additional photon. As a consequence, an electron is excited to the conduction band and via an Auger process, a free electron is created, while the center converts to NV$^0$ \cite{Siyushev2013}. NV$^0$ centers display a ZPL at \unit[575]{nm}, thus, light with wavelengths shorter than \unit[575]{nm} is needed to excite NV$^0$. As a consequence, light with wavelengths between \unit[637]{nm} and \unit[575]{nm} only excites NV$^-$ but not NV$^0$. This can be used to detect the charge state of NV centers \cite{Shields2015, Doi2016} even without changing the charge state: only if the center is NV$^-$, under excitation with e.g.\ \unit[594]{nm} light PL occurs, while no PL is observed in the NV$^0$ case.  To convert NV$^0$ back to NV$^-$ it is necessary to excite NV$^0$: If an electron from the lower orbitals of NV$^0$ is excited, an electron from deep lying orbitals in the valence band can be promoted to the NV$^0$ and re-charge the center to NV$^-$ \cite{Siyushev2013}. It should be noted that the NV charge state in an ensemble can strongly depend on previous laser excitation of the NVs \cite{Dhomkar2018}.

\subsection{Sensitivity \label{sec:sensitivity}}
Sensing using NV centers can be classified into three broad categories, considering the physical quantity to be detected and the degree of freedom of the NV used:
\begin{itemize}
	\item Detection of the (electronic) spin resonance frequencies (magnetometry, temperature sensing, conductivity measurement)
	\item Detection of NV PL intensity and excited state lifetime (near-field energy transfer, Förster-Resonance Energy Transfer (FRET) imaging, near field sensing)
	\item Detection of the charge state of NV centers (sensing of electrostatic environment) 
\end{itemize}

Measurements based on the spin degrees of freedom are most mature, while near-field based sensing and using the NV charge state for sensing are rather novel. Regarding electronic spin-based sensing, we introduce the sensitivity $\eta$ for DC magnetometry \cite{Rondin2014}:
\beq\label{eq:sens_mag}
\eta=\frac{h}{g\mu_B}\frac{\Delta\nu}{\sqrt{I_0}C} \label{eq:sensitivity}
\eeq
where ${I_0}$ is the NV PL rate, $C$ is the ODMR contrast,  $\Delta\nu$ the linewidth of the ODMR resonance, $g$ the Land\'e factor and  $\mu_B$ the Bohr magneton.
Considering equation (\ref{eq:sens_mag}), improving sensitivity can be obtained by:
\begin{itemize}
	\item Increasing ${I_0}$ via increasing the collected photon rates or the collection efficiency for NV PL \cite{Rondin2014, Clevenson2015, lee2011planar,  Riedel2014, li2015efficient, wan2018efficient, Neu2014, Ishiwata2017, Michl2014}
	\item Decreasing $\Delta\nu$, which is connected to increasing the coherence time T$_2^{\star}$  \cite{de2017tailoring, ohno2012engineering, Chandran2016, lovchinsky2016nuclear, yamamoto2013extending, tetienne2018spin, Oliveira2015}
	\item Increasing the efficiency of the spin readout (in Eq.\ \ref{eq:sensitivity} represented via $C$)  \cite{Wolf2015, hrubesch2017efficient, Chatzidrosos2017, jensen2014cavity, Shields2015, Ariyaratne2018, lovchinsky2016nuclear,el2017optimised,shin2012room}.
\end{itemize}

Equation (\ref{eq:sens_mag}) is valid for a single NV, if we use an ensemble including $N$ NV centers, the sensitivity improves by a factor of $\sqrt N$. Indeed, to date,  optimal sensitivities in the pt/$\sqrt{\mathrm{Hz}}$ are reached using ensembles of $N\sim10^{11}$ NVs \cite{wolf2015subpicotesla}. For a recent review focusing on enhancing the sensitivity of NV ensemble-based sensors see Ref.\ \cite{Barry2019}.

For sensing schemes based on the NV charge state and FRET-based sensing and imaging, it is much more challenging to define a sensitivity. However, due to optical read-out, enhancing photon collection will enhance sensing capabilities. For FRET-based sensing, the quantum efficiency QE of the NV transitions, that is the probability for a radiative transition between excited and ground state, also plays a major role. Recent work indicates that the QE for shallow NVs, required for FRET-based sensing (\unit[4.5]{nm} below surface), amounts to 0.7 while defects deeper in the diamond reach QE of 0.86 \cite{Radko2016}. 

In the following, we review the above mentioned methods and the main parameters influencing their sensitivity.

\subsubsection{Collection efficiency \label{sec:colleff}}
As introduced in Section \ref{sec:NVbasics}, the PL of NV centers originates in a semiclassically description from two emitting dipoles in the plane orthogonal to the NV axis.
Two main factors limit the PL collection from NV centers in SCD: 
\begin{itemize}
	\item Refraction effects at the SCD-air interface. Due to the high refractive index of diamond (2.4), a significant fraction of the PL undergoes total internal reflection limiting the collected fraction of PL often to a few percent (see also Section \ref{sec:nanostructuring}). 
	\item Dipole orientation and directionality of NV radiation: Optimal collection of PL is obtained for dipoles parallel to the SCD surface \cite{Neu2014}. However, the most common CVD SCDs have (100) surfaces. Consequently, the NV axis lies under an oblique angle (54.7$^\circ$) to the SCD surface, leaving the NV's dipoles in a non-optimal orientation. Consequently, a fraction of PL even if it escapes the SCD might not be collected because it is emitted into angles that exceed the numerical aperture of the collection optics. 
\end{itemize}

Optimal orientation of NV dipoles can be obtained using SCD with (111)-oriented surfaces \cite{Neu2014}. (111)-oriented CVD growth can even provide NV centers that all align along one [111] direction, namely the growth direction of the crystal \cite{Ishiwata2017,Michl2014,Lesik2014}. No NVs in other equivalent directions are formed and a perfectly aligned ensemble can be created. 
As (111)-oriented CVD growth is technically very challenging, researchers also investigate similar growth directions like (113) that allow for an almost optimal orientation of NV centers \cite{Lesik2015}.  

Regarding refraction effects, PL collection can be improved further via shaping SCD into nanostructures as reviewed in Section \ref{sec:nanostructuring}. 

\subsubsection{Coherence time \label{se:coherencetime}}
Shallow NVs typically used for sensing, present degraded coherence properties compared to NVs well below the surface \cite{OforiOkai2012, Romach2015}: Their coherence times T$_2$ and T$_2^{\star}$ reduce, leading, among other drawbacks, to an increased ODMR linewidth $\Delta\nu$ and reduced sensitivity for DC magnetic fields (see Eq.\ \ref{eq:sensitivity}). The main sources of noise leading to the reduced coherence are:
\begin{itemize}
	\item Surface modified phononic-coupling \cite{Romach2015}
	\item Surface electronic spin bath \cite{Romach2015, kim2015decoherence}
	\item Paramagnetic defects due to implantation \cite{yamamoto2013extending, tetienne2018spin}
\end{itemize}
Methods to improve coherence in shallow NVs are e.g. reviewed in Ref.\ \cite{Bernardi2017}. 

As shallow NVs are often created using low energy (keV) ion implantation, noise due to the implantation process is crucial. Paramagnetic defects responsible for magnetic noise are vacancy clusters or chains created during post-implantation annealing of the diamond \cite{yamamoto2013extending,de2017tailoring}. Ref.\ \cite{de2017tailoring} identifies the neutral di-vacancy complex $V_2$ as the main source of decoherence and avoids its formation via charging the vacancies in the space charge larger of a p-i-junction using boron-doped diamond \cite{de2017tailoring}.  Other work employs annealing at elevated temperatures to anneal out vacancy complexes \cite{yamamoto2013extending, tetienne2018spin}. Alternatively, shallow NV centers can be produced using the $\delta$-doping technique (see also Section \ref{sec:diamondgrowth}). The subsequent creation of vacancies e.g.\ using a focused electron beam followed by annealing, introduces in general lower levels of magnetic noise and enhanced coherence times have been demonstrated \cite{McLellan2016}.

The noise due to the surface of the diamond strongly depends on the pre-treatment of the surface as many functional groups can be present on the surface \cite{Krueger2007} as well as e.g.\ damage due to plasma treatment.  Improvement of single NV coherence was observed after wet oxidizing chemistry and annealing in oxygen environment \cite{lovchinsky2016nuclear}, however, other groups were not able to reproduce the enhancement for shallower NVs \cite{Yamano2017}. Considering plasma treatments, Ref.\ \cite{Oliveira2015} shows increased coherence times for NVs below surfaces that have been treated using an oxygen plasma with low acceleration of ions towards the etched diamond (low-bias, soft plasma).

\subsubsection{Readout and charge state \label{sec:readout}}
For most sensing experiments, PL-based read-out of the NV electronic spin is performed (ODMR, see Section \ref{sec:NVbasics}). Though being technically simple and versatile, this method has its limitations in the fact that the read-out laser re-polarizes the spin to the m$_s$=0 state and low collection efficiencies render single-shot read-out of the electronic spin impossible at room temperature. 

Improving ODMR-based spin read-out is feasible using:
\begin{itemize}
	\item Repetitive readout \cite{lovchinsky2016nuclear}
	\item Phase sensitive (lock-in) methods \cite{el2017optimised,shin2012room}
	\item Noise reduction by mutual referencing of discrete readout windows \cite{Wolf2015}
\end{itemize}

For a repetitive read-out, the NV electronic spin state is transferred into the nuclear spin state of \textsuperscript{14}N. In contrast to the electronic spin, the nuclear spin can be repeatedly probed via quantum non demolition measurements \cite{neumann2010single}. The advantage here is that the NV state is not reset to $m_s=0$ during readout. Ref.\ \cite{el2017optimised} presents a technique modulating the microwave frequency used for spin manipulation  or of the amplitude of static magnetic fields; subsequently demodulating the PL response using  lock-in amplification suppresses noise at low frequency (typically below \unit[30]{kHz}). Another way to reduce noise due to uncertainties in microwave frequency or power is to subtract PL signals retrieved during subsequent pulse repetitions \cite{Wolf2015}. Using this approach, microwave noise with a correlation time exceeding the duration of a single pulse sequence is filtered out. 

In recent years, novel approaches to read out NV electronic spins beyond ODMR arose, which we summarize here briefly. For a more detailed review see Ref.\ \cite{Hopper2018}. Novel read-out methods that partly bypass the challenge of low collection efficiency include:
\begin{itemize}
	\item  Absorption of infrared (IR) light due to transitions within the NV singlet system 
	\cite{Chatzidrosos2017,jensen2014cavity}
	\item{Laser threshold-related magnetometry \cite{Jeske2016,Dumeige2019}}
	\item Spin to charge conversion for more efficient spin-readout \cite{Shields2015,Ariyaratne2018}
	\item Readout based on the detection of photo-currents \cite{siyushev2019}
\end{itemize}

The absorption via the singlet transition ${}^{1}$E$\to$ ${}^{1}$A$_1$  (level scheme see Section \ref{sec:NVbasics}) depends on the population of the involved levels. As these levels are mainly populated from the $m_s=\pm1$ states in the triplet (assuming the NV is pumped e.g.\ using green laser light), the IR absorption monitors the spin resonances in the triplet and consequently allows the measurement of magnetic fields \cite{Chatzidrosos2017,jensen2014cavity}.  To measure the weak absorption on the ${}^{1}$E$\to$ ${}^{1}$A$_1$ transition, it is advantageous to enclose the diamond in a cavity or at least to use a multipass configuration to enhance the signal. Using miniaturizes Fabry Perot cavities sensitivities down to \unit[28]{pt/$\sqrt{\mathrm{Hz}}$} have been shown \cite{Chatzidrosos2017}.

Another approach to avoid the bottleneck of low PL detection  is laser threshold magnetometry, which promises sensitivities down to the ft/$\sqrt{\mathrm{Hz}}$ regime \cite{Jeske2016}. In this approach, NV centers serve as gain medium for a laser. A change in spin state and thus in PL intensity translates into switching on and off the laser, which operates very close to the laser threshold. However, this requires stimulated emission and population inversion in the NV triplet levels. Despite the fact that stimulated emission has been observed \cite{Jeske2017}, ionization dynamics counteracting stimulated emission render this approach challenging. Combining the previously mentioned approaches, Ref.\ \cite{Dumeige2019} presents a hybrid laser threshold magnetometer. The laser gain is supplied by an external laser material lasing at the ${}^{1}$E$\to$ ${}^{1}$A$_1$ transitions wavelength. The diamond is inserted into the cavity of the external laser. Enhanced absorption on the IR transition increases the losses for the laser and stops laser emission. 

In spin-to-charge conversion, the spin state of the NV center is mapped to a charge state (NV$^-$ or NV$^0$) \cite{Shields2015,Ariyaratne2018}. In brief, the technique is based on the fact that NV$^-$ centers are easily ionized from the triplet excited state ${}^{3}$E. In contrast, a   transition to the singlet, which is more likely to occur if the center is in the $m_s=\pm1$ state, protects the center from ionization. Consequently, $m_s=0$ is mapped to NV$^0$ and $m_s=\pm1$ is mapped to NV$^-$. Following the ionization, the charge state is read-out as described in Section \ref{sec:NVbasics}. Using this read-out method, enhanced sensing with a 25-fold increase in imaging speed has been demonstrated \cite{Ariyaratne2018}.

Each cycle of ionization and re-charging of a NV center under optical illumination creates an electron in the conduction band which can be detected in the form of a photocurrent if an electric field is applied using electrodes on the SCD surface. Photocurrents in the pA are observed for a single NV center and the contrast of the photoelectrical read-out is comparable to the ODMR based read-out \cite{siyushev2019}.  

\subsubsection{Ensemble density and alignment of NVs for wide field imaging \label{sec:ensdens}}
In wide field magnetic imaging, a thin layer of NV centers is used. To enable imaging a larger sample area, NV PL is collected using a camera. Typically, hundreds  to 10000 of NV centers contribute to the signal in every pixel \cite{steinert2010high,Chipaux2015}. In this way, every pixel encodes the properties of the local magnetic field. 

Two properties of the NV layer determine the magnetic field sensitivity:
\begin{itemize}
	\item The density of NVs 
	\item  The alignment of NVs
\end{itemize}

The most common way to increase the NV density is to increase the nitrogen implantation dose. This has two effects on sensitivity: On the one hand, it increases the PL rate and thus improves the sensitivity scaling with $\sqrt{N}$, where $N$ is the number of NV defects. On the other hand, it increases the density of implantation-related defects (see also Section \ref{se:coherencetime}) and decreases the coherence \cite{pham2011magnetic}. These two counteracting effects have been found to results in a sensitivity which can be independent of the implantation dose \cite{tetienne2018spin}. The spin coherence time $T_2$ influences the range of AC frequencies to which the NV center is sensitive \cite{taylor2008high}. Consequently, the optimal NV density is determined by the specific needs of each sensing application.  For a recent review focusing on the sensitivity using NV ensembles see Ref.\ \cite{Barry2019}

For wide field imaging, also the orientation of NV centers has to be taken into account. NV centers oriented along different equivalent directions  sense the projected field onto their high symmetry axis. Consequently, for an ensemble and an arbitrarily aligned magnetic field, 4 pairs of ODMR resonances are observed. Using these 4 ODMR resonance pairs, the magnetic field vector can be reconstructed. However, the ODMR contrast for each of the resonances is strongly reduced as only 25\% of the NV centers contribute to each resonance line. For high-sensitivity experiments, consequently the use of an NV ensemble that is aligned along only one of the equivalent directions is advantageous. Using diamond oriented in $[111]$ direction and creating NV centers during growth, perfectly aligned ensembles with high density ($d=10^{19}$ cm\textsuperscript{-3}) and  high ODMR contrast, comparable to the one reported for single NVs, have been demonstrated \cite{Ishiwata2017}. 

\section{Spatial resolution in NV sensing \label{sec_reso}}
Using ensembles of NV centers often involves wide field imaging, where a larger diamond area is illuminated and NV PL is detected using a camera \cite{Shao2016,LeSage2013,Nowodzinski2015,Schlussel2018,Horsley2018}. This technique typically leads to a spatial resolution on the order of \unit[1]{$\mu$m} \cite{Shao2016,LeSage2013,Nowodzinski2015,Horsley2018}. This spatial resolution allows e.g.\ to investigate microwave devices \cite{Horsley2018}, integrated circuits \cite{Nowodzinski2015} or magnetic features in cells \cite{LeSage2013}. In wide field imaging, the signal on each pixel of the camera involves hundreds to 10000 NV centers \cite{steinert2010high,Chipaux2015}. DC magnetic field sensitivities in the order of $\mu T/\sqrt{Hz}$ have been quoted \cite{Chipaux2015}. Enhancing the sensitivity can be obtained via increasing the number of NV centers participating in the signal creation. However, as discussed in Section \ref{sec:ensdens}, enhancing the volume density of NVs in SCD does not necessarily enhance the sensitivity. An alternative route enhances the number of NV centers via increasing the effective SCD volume contributing to the sensing signal: Ref.\ \cite{Clevenson2015} employs light trapping in a SCD that forms a multi-pass waveguide (SCD dimensions \unit[3]{mm} x \unit[3]{mm} x \unit[0.3]{mm} to increase the number of interacting NV centers). Ref.\ \cite{Chatzidrosos2017} uses NV infrared absorption in a cavity realizing a probed volume of \unit[390]{$\mu$m} x \unit[4500]{$\mu$m$^2$}. Both approaches lead to sensitivities in the order of \unit[10]{$pT/\sqrt{Hz}$} range.  Ref.\ \cite{Barry2016} demonstrates  \unit[15]{$pT/\sqrt{Hz}$} in a volume of \unit[13]{$\mu$m} x \unit[200]{$\mu$m} x \unit[2000]{$\mu$m}, while Ref.\ \cite{Wolf2015} extends sensitivity even to the sub-$pT/\sqrt{Hz}$  using optimized diamond material and optimized pulse sequences. 

Using a single NV center for sensing, the probed volume is in principle given by the spatial extent of the electronic wave-function of the NV center. Recent simulations show that the electrons bound to the NV defect are closely localized within a few lattice constants \cite{bockstedte_ab_2018} rendering the NV center a truly atomic-scale sensor. However, the spatial resolution will be limited by the distance of the sensing NV center to the sample under investigation which is ultimately limited by the distance of the NV center to the diamond surface. NV centers have been observed to be stable \unit[2]{nm} below a SCD surface \cite{Loretz2014}. In addition to this distance to the SCD surface, technical limitations of the scanning probe imaging have to be considered which lead to an additional stand-off distance: Using SCD scanning probes, NV-to-sample distances around \unit[25]{nm} \cite{Appel2015, Maletinsky2012} at ambient conditions and distances around \unit[10]{nm} in vacuum at cryogenic temperature \cite{Thiel2016} have been demonstrated yielding spatial resolutions in the same order of magnitude \cite{Maletinsky2012}.  Typically, scanning probe microscopy using single NV centers reaches $\mu T/\sqrt{Hz}$ sensitivity for DC magnetic fields, while sensitivities on the order of $nT/\sqrt{Hz}$ have been shown for AC magnetometry \cite{Appel2015, Maletinsky2012}. While these sensitivities are significantly lower than obtained using ensemble magnetometry, one has to take into account that the color center is placed in the high field region of the sample. Considering that e.g.\ dipolar magnetic fields decay with the third inverse power of the distance to the sample \cite{Rondin2014}, the lower sensitivity is at least partially compensated by the close proximity of single NVs to the sample in a scanning probe geometry.  

\section{Selected sensing/imaging applications for each type of diamond \label{sec_applications}}
Applications of color centers in diamond have been summarized in several recent reviews focusing on magnetometry \cite{Rondin2014, Bernardi2017}, NV sensing as a probe for solid state physics \cite{casola2018}, single spin magnetic resonance \cite{Wrachtrup2016} and the coupling of spins and mechanical oscillators \cite{lee_topical_2017}.  In the following section, we select a few recent and novel applications using different diamond types to illustrate the points discussed above. 

In addition to demonstrating novel applications, a trend has been arising to miniaturize NV-based sensing devices to enable mobile use and potential industrial applications \cite{Stuerner2019}: via connecting the diamond to micro-optics (graded index lenses), integrating microwave driving on printed circuit boards and using an LED instead of a laser for excitation, the volume of a diamond sensing device has been reduced to less than \unit[3]{cm$^3$}. This small footprint in combination with a low power consumption of \unit[1.5]{W} allows for highly-flexible field-use of the diamond device opening the route toward novel applications. Additionally, recent work presents the first integration of diamond sensors into  complementary metal-oxide-semiconductor (CMOS) technology, paving the way toward chip-scale diamond quantum sensors \cite{Kim2018}. 
 
\subsection{Single-crystal, bulk diamond and F\"orster Resonance Energy Transfer (FRET): novel approaches to NV sensing to enhance versatility of NV sensors}
While NVs in high-purity, single-crystal, bulk diamond (SCD) typically have superior properties compared to NVs in often less pure NDs, probing sample dynamics with high spatial resolution requires to transfer the sample under investigation onto the SCD surface and spatial resolution is typically limited by diffraction to around \unit[0.5]{$\mu$m}. Nevertheless, shallow NVs in SCD have been used for various applications probing  micron-scale dynamics especially in the context of novel materials and current imaging. The almost back-action-free detection of currents in graphene, where currents in the $\mu$Ampere regime are measured with a resolution of \unit[0.5]{$\mu$m} \cite{Tetienne2017graphene}, and the usage of  NV centers to investigate failure of integrated circuits \cite{Nowodzinski2015} illustrate their potential to analyze currents on the micron scale. Very recently, NV centers measured the band bending inside a diode like device \textit{in-situ} \cite{broadway2018bandbending}. Whereas in Ref.\ \cite{Lovchinsky2017} shallow NV centers are used as sensor to perform nuclear magnetic resonance spectroscopy of atomically-thin hexagonal boron nitride layers. Thus NV centers allow to bring the technology of magnetic resonance truly to the nanoscale. For probing liquid samples, Kehayias et al.\ structured SCD to form a microfluidic chip and flow the substance under investigation through the diamond device \cite{Kehayias2017microfluidics}. 

In contrast to the above mentioned applications which all rely on the spin-based sensing capability of NVs, sensing relying on the dipolar nature of the NV center is less advanced. There have been first demonstrations using nearfield-based energy transfer (F\"orster Resonance Energy Transfer, FRET) to detect dye molecules attached to the surface of NDs \cite{Tisler2011, Mohan2010a}. Additionally, a scanning ND attached to an AFM tip has been used to image graphene flakes with nanoscale resolution \cite{Tisler2013a}. Other authors reported non-successful attempts to establish FRET using NDs potentially due to a lack of control over the surface properties of NDs \cite{Sekatskii2015}. Consequently, it would be highly-advantageous to broaden FRET-based sensing to shallow color centers in SCD. For sensing and imaging approaches using the dipolar nature of NV centers, it is significant to consider the role of the two electric transition dipoles, $\vec{X}$ and $\vec{Y}$ as we discuss in detail in section \ref{sec:append}. As a first step toward FRET-based imaging using color centers in SCD, we here present quenching due to graphene deposited onto the SCD surface.   

\begin{figure}
	\centering
	\includegraphics[width=0.8\linewidth]{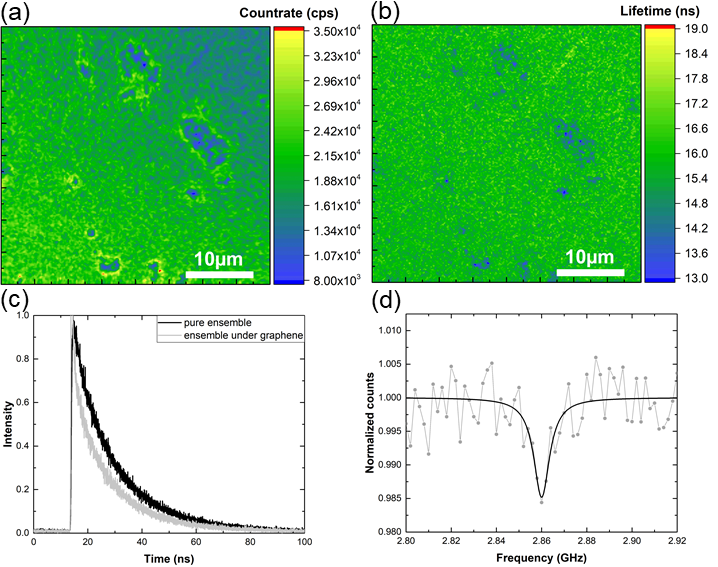}
	\caption{(a) PL map of the NV ensemble with graphene agglomerates (darker area) on the SCD surface (PL detected at wavelengths above \unit[650]{nm}, excitation \unit[0.7]{mW} at \unit[532]{nm}. (b) Lifetime map of the same area like in (a). The excited state PL lifetime $\tau$ decreases up to $4\,$ns under the agglomerates. (c) NV ensemble lifetime measurement under the graphene (grey, $\tau$=\unit[14.5]{ns}) and next to a graphene agglomerate (black, $\tau$=\unit[16.4]{ns}). (d) ODMR measurement of the quenched NV centers under a graphene agglomerate. \label{fig:graphene}}	
\end{figure}

We use commercial, high-purity, chemical vapor deposited, electronic grade SCD from Element Six ([N]$^s$<\unit[5]{ppb}, B<\unit[1]{ppb}). To create shallow NVs, we implant the sample with $6\times10^{11}\,$cm$^{-2}$ nitrogen ions (Innovion, USA). Via annealing to \unit[800]{$^{\circ}$C} in vacuum and cleaning in boiling acids (nitric, sulfuric, perchloric acid, 1:1:1), we create a layer of NV centers on average \unit[7]{nm} below the SCD surface. From this clean surface, we observe homogeneous NV PL. We estimate the number of NV centers in the laser focus of our confocal microscope to be on the order of < 10. The NV ensemble shows a spatially consistent lifetime of \unit[15.5$\pm$1]{ns}, which is slightly higher than the bulk lifetime of \unit[12.9$\pm$0.1]{ns} \cite{Collins1983} due to a decrease in effective refractive index close to the surface. We now apply graphene flakes from solution (pristine graphene monolayer flakes from graphene supermarket with an average flake size of \unit[550]{nm} (\unit[150-3000]{nm}), dispersed in ethanol). We spin coat this solution as-received with \unit[600]{rpm} onto the SCD surface. We observe a slight trend to agglomeration of graphene. While in principle not desired, these agglomerates enable to straightforwardly observe the interaction of the NV centers with the graphene: We identify agglomerates via white light illumination of the SCD surface as well as Raman spectroscopy revealing a G Band at \unit[1586]{cm$^{-1}$}. At the position of graphene agglomerates, we observe a reduced PL intensity [see Fig.\ \ref{fig:graphene}(a)] as well as a clearly decreased NV lifetime [see Fig.\ \ref{fig:graphene}(b)]. We check our results on several positions on the sample and find typical results as shown in Fig.\ \ref{fig:graphene}(a)-(c). From the decrease in lifetime and the graphene F\"orster radius of \unit[15]{nm} \cite{Tisler2013a}, we estimate the average distance of the graphene to the NV centers to be \unit[19]{nm} which is higher than the average depth of the NV centers of \unit[7]{nm}. We attribute this finding to a strong quenching of NV centers very close to the surface. These centers consequently do not contribute significantly to the measured PL signal which is dominated by the NVs deeper inside the diamond. Using the distance to the graphene, we estimate a reduction to 75\% of the non-quenched PL. We observe a stronger reduction to 50\%. However, our simplified consideration does not take into account the unknown absorption of the excitation laser as well as the NV PL due to the graphene agglomerate. We now test if NV ensembles below graphene agglomerates still show the typical spin properties of NV centers. To this end, we record an ODMR of an NV ensemble [see Fig.\ \ref{fig:graphene}(d)] below one of the agglomerates, which shows the characteristic ODMR resonance. This result thus strongly indicates that NV centers can serve as multifunctional sensors, which probe the presence of other dipoles as well as the magnetic environment simultaneously. 

\subsection{Nanodiamonds for sensing and imaging}
NDs are versatile tools in the life sciences as they can enter living cells and organisms. Their applications do not only cover sensing and imaging but also extend to drug delivery and tissue engineering.  For recent reviews on these applications see Refs. \cite{Claveau2018BioRev,Schirhagl2013,chipaux2018nanodiamonds,chen2017diamond,prabhakar2019nanodiamonds,Laan2018nanodiamonds,wu2016quantumdevicesbio,Hsiao2016}. Color centers in NDs first serve as non-bleaching fluorophores and additionally sense temperatures and magnetic fields inside cells. Moreover, novel approaches using the NV center's charge state as a resource allow to sense the electrostatic environment of the NV center \cite{Karaveli2016,Petrakova2015} and potential changes as small as \unit[20]{mV} have been measured using this technique. 

\begin{figure}
	\centering
	\includegraphics[width=1\linewidth]{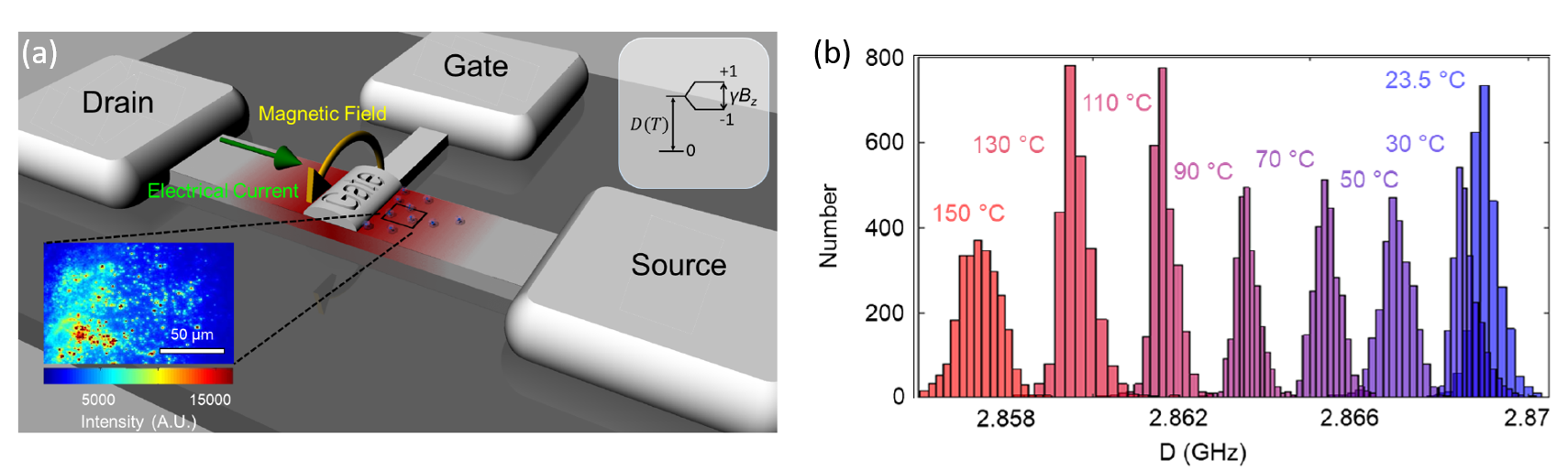}
	\caption{(a) Schematic of temperature and current (magnetic field) imaging using ND coated onto a transistor. (b) Zero-field-splitting statstics of NDs with NV ensembles. Reprinted from Ref.\ \cite{Foy2019}. \label{fig:NDtransistor}}	
\end{figure}
Considering electronic devices, NDs with NV ensembles enable the simultaneous measurement of local temperature and electric current distribution \cite{Foy2019}. These quantities are relevant in the context of high electron mobility electronics, where local heating occurs and potentially induces device degradation [see Fig.\ \ref{fig:NDtransistor}(a)]. Both parameters can be measured in parallel as the local temperature affects the NV's zero-field-splitting, while the local current density influences the splitting of the m$_s$=-1 and m$_s$=1 states as a result of the local magnetic field. Despite a certain scatter in the zero-field-splitting of individual NDs [see Fig.\ \ref{fig:NDtransistor} (b)], the relevant temperature increases can be measured. For this applications, the ability to deposit NDs on almost all desired substrates and devices is highly advantageous.

\subsection{Nanostructured diamond: pushing spatial resolution to the limit}
Nanostructured SCD especially in the form of scanning diamond pillars has been used to image magnetic features with nanoscale resolution revealing magnetic features like superconducting vortices \cite{Thiel2016}, skyrmions \cite{Jenkins2018, Gross2018} and magnetic properties in antiferromagnets \cite{Gross2017}. Moreover, scanning NVs locally imaged the magnetic field arising due to microwave currents in a stripline \cite{Appel2015}. As a novel field of application, NV centers in scanning tips have been used to probe the local conductivity of metallic structures \cite{Ariyaratne2018} and have proven to be useful for the investigation of the magnetic field of a hard-disk write head \cite{Jakobi2017}. To avoid the need for scanning pillars, also the sample under investigation e.g.\ a cell can be scanned over a stationary pillar. Using this approach, clusters of iron-storage molecules (Ferritin) have been imaged in a single cell \cite{Wang2019}.
\begin{figure}
	\centering
	\includegraphics[width=0.6\linewidth]{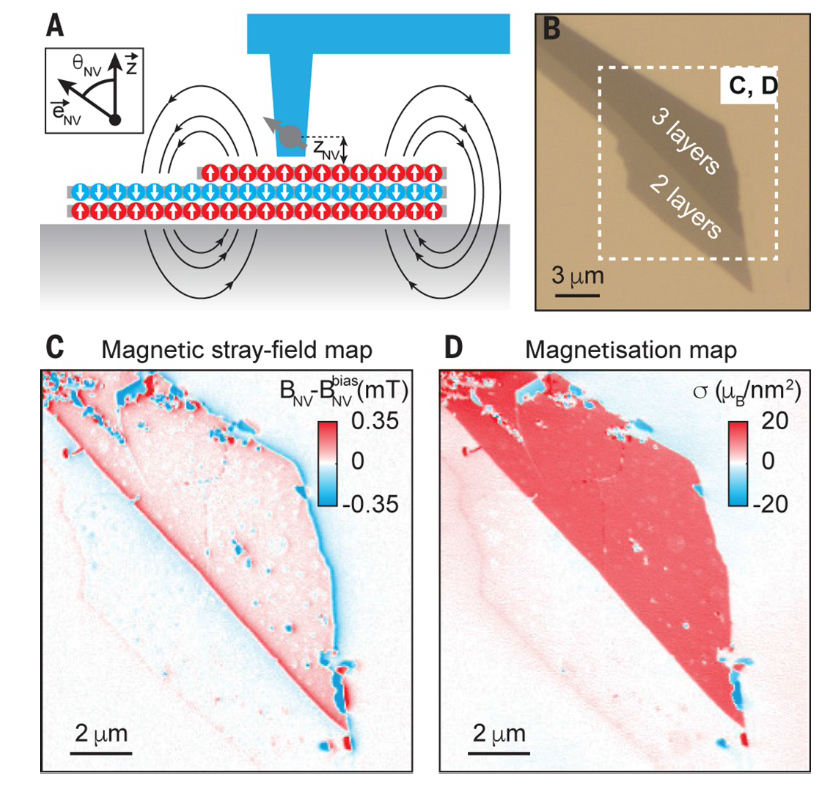}
	\caption{(a) Schematics of single NV scanning magnetometry (b) Optical micrograph of the CrI$_3$ flake with regions with 2 and 3 layers. (c) Magnetic stray field as imaged using a single NV center. (d) Calculated magnetization of the flake. Reprinted with permission from Ref.\ \cite{Thiel2019}.  \label{fig:2Dmagn}}	
\end{figure}

Recently, the discovery of ferromagnetic behavior in two-dimensional van der Waals  materials has triggered significant research interest. An example for such a material is CrI$_3$. Using scanning NV magnetometry [see Fig.\ \ref{fig:2Dmagn}(a)], fundamental questions on the magnetism in CrI$_3$ have been addressed \cite{Thiel2019}: The measurements show that flakes with an even number of layers show no resulting magnetization in contrast to flakes with an odd number of layers [see Fig.\ \ref{fig:2Dmagn}(b)-(d)]. Due to the high spatial resolution and the low back-action in this sensing approach, it is possible to observe direct connections between structural and nanoscale magnetic properties of the material. Consequently, magnetic imaging using a scanning single NV center can help to understand magnetic two-dimensional materials which are candidates for future spintronic devices. 

\section{Conclusion and Outlook}
In recent years, sensing using color centers has been applied to many novel fields of applications. Novel sensing and imaging modes are under development taking into account e.g.\ near-field interactions of color centers. Using such near field interactions, color centers will be efficient probes for the presence of other light emitting systems like e.g.\ quantum dots, molecules or two-dimensional materials \cite{nelz2019near}. Novel approaches to read-out the spin state of NV centers have made magnetic sensing faster and more efficient. Sensing capabilities of color centers have been extended not only to spin-based temperature sensing \cite{Kucsko2013}, but also to all optical temperature sensing \cite{Ngyuen2018}. The controlled coupling of color centers to strain and thus vibrational modes in nanoscale devices opens the route towards new control schemes in sensing and quantum information \cite{Barfuss2015}. Moreover, the charge state of NV centers close to diamond surfaces has been found to depend strongly on the treatment of the surfaces \cite{Rondin2010}. Consequently, charge-state changes might be used as a future sensing resource. These ongoing developments will render color centers truly multi-functional sensors and the full potential of combining these recent results in sensing is still mainly un-explored.

Developments of new materials are ongoing and for example manufacturing of smaller and smaller nanodiamonds with active color centers will be pave the way towards enhanced usage of nanodiamonds inside cells \cite{bolshedvorskii2019single}.  Integrated sensors pave the way towards industrial applications and color center sensors have enabled new insights into the nanoworld e.g.\ in the investigation of novel two-dimensional magnetic materials \cite{Thiel2019}.

Challenges still arise from decoherence due to surfaces as well as implantation-related effects \cite{vanDam2019}, and improvements and deeper understanding of the underlying effects could trigger a significant enhancement in NV sensor performance. A very recent example of novel approaches towards mitigating negative effects of nanofabrication is Ref.\ \cite{ruf2019optically} in which a combination of electron-irradiation, high-temperature annealing and optimized etching lead to the creation of optically coherent NV centers. Similar approaches might be applied to sensing devices, enhancing their performance.   
\section{Appendix \label{sec:append}}
\subsubsection{Influence of the orientation of NV centers on near-field based sensing \label{sec:dipoleor}}
For sensing and imaging approaches using the dipolar nature of NV centers, it is significant to consider the role of the two electric transition dipoles, $\vec{X}$ and $\vec{Y}$. The detection of near-field processes typically involves measuring the lifetime of the interacting systems. The following calculation illustrates how the observed lifetime is influenced by the existence of two dipoles  $\vec{X}$ and $\vec{Y}$.

$\vec{X}$ and $\vec{Y}$ are orthogonal to each other and located in the plane perpendicular to the NV axis (see Section \ref{sec:NVbasics}, \cite{davies1976optical, epstein2005anisotropic}), while local strain determines the in plane orientation $\vec{X}$ and $\vec{Y}$. 
Without local strain and the NV axis along  [111], $\vec{X}$ and $\vec{Y}$ orient as $X\|[\bar{1}\bar{1}2]$ and $Y\|[1\bar{1}0]$ \cite{epstein2005anisotropic}.

Following a semiclassical approximation, the radiative decay rate $ \mathbf{\Gamma}^{rad}$ associated to a transition from an excited state to a ground state is given by  \cite{davies1976optical}

\beq
\mathbf{\Gamma}^{rad}=\frac{\mathcal{P}}{\hbar\omega} \label{eqn_gamma_gen}.
\eeq

Here, $\mathcal{P}$ is the power dissipated by the transition dipole $\vec{\mu}$. The lifetime $\tau$ for this transition is given as the inverse of the radiative decay rate $ \mathbf{\Gamma}^{rad}$ (assuming a quantum efficiency of unity). 

In a homogeneous medium with a dielectric constant $\epsilon_1$, we can write:
\beq
\mathbf{\Gamma}^{rad}_{hom}=\frac{\mathcal{P}_{hom}}{\hbar\omega}, \qquad \mathcal{P}_{hom}(\vec{\mu}^2)= \vec{\mu}^2\sqrt{\epsilon_1}\omega^{4}/3c^3. \label{eqn_lay_hom}
\eeq

Assuming a homogenous dielectric is suitable for an NV center buried deeply in diamond. In this case, the observed decay rate (and hence the lifetime) will be given by the average of $\mathbf{\Gamma}^{\vec{X},rad}_{hom}$ and $\mathbf{\Gamma}^{\vec{Y},rad}_{hom}$ :

\beq
\label{eq_gamma_tot_X_Y_hom}
\mathbf{\Gamma}^{rad}_{hom}=\frac{1}{2}(\mathbf{\Gamma}^{\vec{X},rad}_{hom}+\mathbf{\Gamma}^{\vec{Y},rad}_{hom}).
\eeq

To model a NV center close to the diamond surface, we have to use a a layered system, with a NV center buried at a depth $h$. We decompose the power dissipated by the two dipoles $\mathcal{P}_{lay}^{\vec{X}}$, $\mathcal{P}_{lay}^{\vec{Y}}$  in components parallel and orthogonal to the surface:

\beq
\label{eqn_lay_X}
\mathcal{P}_{lay}^{\vec{X}}=\frac{{\vec{\mu}_{\perp,\vec{X}}}^2}{\vec{\mu}_{\vec{X}}^2} \mathcal{P}_{lay,\perp}+\frac{{\vec{\mu}_{\parallel,\vec{X}}}^2}{{\vec{\mu}_{\vec{X}}^2}} \mathcal{P}_{lay,\parallel}\stackrel{!}{=}{a_{\perp}^{\vec{X}}}^2\mathcal{P}_{lay,\perp} + {a_{\parallel}^{\vec{X}}}^2\mathcal{P}_{lay,\parallel},
\eeq

\beq
\label{eqn_lay_Y}
\mathcal{P}_{lay}^{\vec{Y}}=\frac{{\vec{\mu}_{\perp,\vec{Y}}}^2}{\vec{\mu}_{\vec{Y}}^2} \mathcal{P}_{lay,\perp}+\frac{{\vec{\mu}_{\parallel,\vec{Y}}}^2}{\vec{\mu}_{\vec{Y}}^2}\mathcal{P}_{lay,\parallel}\stackrel{!}{=}{a_{\perp}^{\vec{Y}}}^2\mathcal{P}_{lay,\perp} + {a_{\parallel}^{\vec{Y}}}^2\mathcal{P}_{lay,\parallel},
\eeq

where $\mathcal{P}_{lay,\perp}$($\mathcal{P}_{lay,\parallel}$) is the power dissipated by the dipole component orthogonal (parallel) to the surface. We note that $\mathcal{P}_{lay,\perp}$ and $\mathcal{P}_{lay,\parallel}$ as well as all deduced quantities depend on the depth $h$ of the NV center. Equations defining quantities and coefficient are highlighted using $\stackrel{!}{=}$.

Considering Eqns.\ \ref{eqn_gamma_gen}, \ref{eqn_lay_hom}, Eq. \ref{eqn_lay_X}, we find the decay rate $\mathbf{\Gamma}^{\vec{X},rad}_{lay}$ for $\vec{X}$:

\beq
\label{eqn_gamma_lay_X}
\mathbf{\Gamma}^{\vec{X},rad}_{lay}=\frac{\mathcal{P}_{lay}^{\vec{X}}}{\hbar\omega}\stackrel{!}{=}F_{\vec{X}}\mathbf{\Gamma}^{rad}_{hom}=({a_{\perp}^{\vec{X}}}^2  F_{\perp}+ {a_{\parallel}^{\vec{X}}}^2  F_{\parallel})\mathbf{\Gamma}^{rad}_{hom},
\eeq

with:

\beq
F_{\perp}\stackrel{!}{=}\frac{\mathcal{P}_{lay,\perp}}{\mathcal{P}_{hom}}\,\,\,\,,\,\,\, F_{\parallel}\stackrel{!}{=}\frac{\mathcal{P}_{lay,\parallel}}{\mathcal{P}_{hom}}.
\eeq

Similarly for $\vec{Y}$

\beq
\label{eqn_gamma_lay_Y}
\mathbf{\Gamma}^{\vec{Y},rad}_{lay}=\frac{\mathcal{P}_{lay}^{\vec{Y}}}{\hbar\omega}\stackrel{!}{=}F_{\vec{Y}}\mathbf{\Gamma}^{rad}_{hom}=({a_{\perp}^{\vec{Y}}}^2  F_{\perp}+ {a_{\parallel}^{\vec{Y}}}^2  F_{\parallel})\mathbf{\Gamma}^{rad}_{hom}.
\eeq

To obtain the total decay rate, we use Eq. \ref{eq_gamma_tot_X_Y_hom}:

\beq
\label{eq_gamma_tot_X_Y}
\mathbf{\Gamma}^{rad}_{lay}=\frac{1}{2}(\mathbf{\Gamma}^{\vec{X},rad}_{lay}+\mathbf{\Gamma}^{\vec{Y},rad}_{lay}).
\eeq

Considering Eq. \ref{eqn_gamma_lay_X} and Eq. \ref{eqn_gamma_lay_Y},

\beq \label{eq_lay}
\mathbf{\Gamma}^{rad}_{lay}=\frac{1}{2}\mathbf{\Gamma}^{rad}_{hom}[   F_{\perp}({a_{\perp}^{\vec{X}}}^2+{a_{\perp}^{\vec{Y}}}^2) +  F_{\parallel}({a_{\parallel}^{\vec{X}}}^2+{a_{\parallel}^{\vec{Y}}}^2) ]=\\
                           \mathbf{\Gamma}^{rad}_{hom}[   a_{\perp}^2F_{\perp}+  a_{\parallel}^2F_{\parallel} ]\\
				           =\mathbf{\Gamma}^{rad}_{hom}F_{lay},
\eeq

with:

\beq
F_{lay}\stackrel{!}{=}a_{\perp}^2F_{\perp}+  a_{\parallel}^2F_{\parallel},
\eeq

and

\beq
a_{\perp}^2\stackrel{!}{=}\frac{{a_{\perp}^{\vec{X}}}^2+{a_{\perp}^{\vec{Y}}}^2 }{2}\,\,\,\,\,,\,\,\,\,a_{\parallel}^2\stackrel{!}{=}\frac{{a_{\parallel}^{\vec{X}}}^2+{a_{\parallel}^{\vec{Y}}}^2}{2}.
\eeq

Considering in detail the possible orientations of  $\vec{X}$ and $\vec{Y}$ in (100)-oriented SCD which is the most common SCD orientation, we find that that $a_{\perp}^2$ and $a_{\parallel}^2$ do not depend on the orientation of $\vec{X}$ and $\vec{Y}$ in the plane orthogonal to the NV axis, nor on which of the equivalent directions the NV occupies and $a_{\perp}^2=\frac{1}{3}$ and $a_{\parallel}^2=\frac{2}{3}$.

Consequently, $ \mathbf{\Gamma}^{rad}_{lay}$ is the same for all equivalent orientations of NV axis and orientations of $\vec{X}$ and $\vec{Y}$. $ \mathbf{\Gamma}^{rad}_{lay}$ and the lifetime of the NV will depend only on the NV depth  $h$. FRET will add a non-radiative channel to the system. 
In this case the total decay rate $\mathbf{\Gamma}_{lay}^{tot}$ for the layered system is:

\beq
\mathbf{\Gamma}_{lay}^{tot}=\mathbf{\Gamma}_{lay}^{rad}+\mathbf{\Gamma}_{lay}^{nr}=\mathbf{\Gamma}^{rad}_{hom}F_{lay}+\mathbf{\Gamma}_{lay}^{nr}.
\eeq

Consequently, we find a unique decay rate $\mathbf{\Gamma}_{lay}^{tot}$ for an ensemble of NV centers at the same depth $h$. Hence, the observed PL decay will be monoexponential characterized by a lifetime:

\beq
\tau=\frac{1}{\mathbf{\Gamma}_{lay}^{tot}}.
\eeq
Consequently, shalllow NV centers in (100) oriented diamond may form highly-suitable sensors for FRET, despite their non-trivial transition dipoles.
\section*{References}
\bibliography{Literaturverzeichnisaktuell,bernardi_rev,References} 
\bibliographystyle{iopart-num}

\end{document}